\documentclass[preprint,tightenlines,superscriptaddress,showpacs]{revtex4}
\usepackage{epsfig}
\usepackage{dcolumn}
\usepackage{bm}
\usepackage{amsmath}
\usepackage{subfigure}
\usepackage{graphicx}
\newcommand{\Y}{$Y(4220)$}
\newcommand{\EE}{e^+e^-}
\newcommand{\BR}{{\cal B}}
\newcommand{\eetopipihc}{e^+ e^- \to \pi^+ \pi^- h_c}

\newcommand{\eetopipijpsi}{e^+ e^- \to \pi^+\pi^-J/\psi}
\newcommand{\eetoddpi}{e^+ e^- \to D^0D^{*-}\pi^+ +c.c.}
\newcommand{\pphc}{\pi^+ \pi^- h_c}

\newcommand{\ppjpsi}{\pi^+\pi^-J/\psi}
\newcommand{\ddpi}{D^0D^{*-}\pi^+ + c.c.}

\begin{document}

\title{\boldmath Resonant parameters of the \Y}

\affiliation{School of Physics and Nuclear Energy Engineering,
Beihang University, Beijing 100191, China} \affiliation{Institute
of High Energy Physics, Chinese Academy of Sciences, Beijing
100049, China} \affiliation{University of Chinese Academy of
Sciences, Beijing 100049, China}

\author{X.~Y.~Gao}
\affiliation{School of Physics and Nuclear Energy Engineering,
Beihang University, Beijing 100191, China}
\author{C.~P.~Shen}
\email{shencp@buaa.edu.cn} \affiliation{School of Physics and
Nuclear Energy Engineering, Beihang University, Beijing 100191,
China}
\author{C.~Z.~Yuan}
\email{yuancz@ihep.ac.cn}\affiliation{Institute of High Energy
Physics, Chinese Academy of Sciences, Beijing 100049, China}
\affiliation{University of Chinese Academy of Sciences, Beijing
100049, China}

\begin{abstract}

The vector charmoniumlike state \Y\ was reported recently in the
cross sections of $e^+e^-\to \pi^+\pi^-h_c$, $\omega \chi_{c0}$,
$\pi^+\pi^- J/\psi$, and $D^0 D^{*-}\pi^+ + c.c.$ measured by the
BESIII experiment. A combined fit is performed to the cross
sections of these four final states to measure the resonant
parameters of the \Y. We determine a mass $M=(4219.6\pm 3.3\pm
5.1)$~MeV/$c^2$ and a total width $\Gamma=(56.0\pm 3.6\pm
6.9)$~MeV for the \Y, where the first uncertainties are
statistical and the second ones systematic. We determine the lower
limit of its leptonic decay width of around 30~eV, which can be
compared with the theoretical expectations of different models. We
also estimate its partial decay width to $\pi\pi J/\psi$ in
different scenarios. These information is essential for the
understanding of the nature of this state.

\end{abstract}


\maketitle

%%%%%%%%%%%%%%%%%%%%%%%%%%%%%%%%%%%%%%%%%%%%%%%%%%%%%%%%%%%%%%%%
%%%%%     Introduction       Part                  %%%%%%%%%%%%%
%%%%%%%%%%%%%%%%%%%%%%%%%%%%%%%%%%%%%%%%%%%%%%%%%%%%%%%%%%%%%%%%

\section{Introduction}

During the last decade, many new states with hidden charm-quark
pair were discovered~\cite{xyz-review}. The number of observed
states in experiments is more than that of the predicted
charmonium states in potential models for the mass above the
$D\bar{D}$ threshold~\cite{barnes} and there are also charged
states which are obviously not charmonium states. These states,
such as the $X(3872)$~\cite{x-3872}, the
$Y(4260)$~\cite{y-4260-1}, and the
$Z_c(3900)$~\cite{y-4260-2-3900-1,zc-3900-2}, are referred to as
charmoniumlike states or $XYZ$ particles~\cite{xyz-review}.

Among these new charmoniumlike states, there are many vector
states with quantum numbers $J^{PC} = 1^{--}$ that are usually
called $Y$ states, like the $Y(4260)$~\cite{y-4260-1}, the
$Y(4360)$~\cite{y-4360-1}, and the
$Y(4660)$~\cite{y-4360-2-4660-1}. The $Y$-states show strong
coupling to hidden-charm final states in contrast to the vector
charmonium states in the same energy region ($\psi(4040)$,
$\psi(4160)$, $\psi(4415)$) which couples dominantly to open-charm
meson pairs. These $Y$ states are good candidates for new types of
exotic particles and stimulated many theoretical interpretations,
including tetraquarks, molecules, hybrids, or
hadrocharmonia~\cite{xyz-review}.

These $Y$ states were observed at $B$ factories with limited
statistics since they are produced from initial state radiation
processes with data collected at around 10.6~GeV in the
bottomonium energy
region~\cite{y-4260-1,y-4360-1,y-4360-2-4660-1}. The high
precision cross section measurements and the study of these states
in different final states in direct $\EE$ annihilation in the
charmonium energy region from the BESIII experiment supply new
insight into their properties.

In 2013, BESIII reported the cross section measurement of
$\eetopipihc$ at 13 center-of-mass (c.m.) energies from 3.9 to
4.2~GeV and found the magnitude of the cross sections is at the
same order as that of $\EE\to \pi^+\pi^-J/\psi$ but with a
different line shape. Although no quantitative results were given,
the resonant structure at around 4.22~GeV/$c^2$ is
obvious~\cite{pipihc-1}. A combined fit to the BESIII data
together with the CLEO-c measurement at
4.17~GeV~\cite{cleoc_pipihc} results in a resonant structure, \Y,
with a mass of $(4216\pm 18)~{\rm MeV}/c^2$ and a width of $(39\pm
32)$~MeV~\cite{y4220_ycz}, different from any of the known $Y$ and
excited $\psi$ states in this mass region~\cite{pdg}.

In 2014, BESIII reported the cross section measurement of $\EE\to
\omega \chi_{c0}$ at 9 c.m. energies from 4.21 to 4.42~GeV. By
assuming the $\omega \chi_{c0}$ signals come from a single
resonance, BESIII reported a resonant structure with the mass and
width of  $(4230\pm 8\pm 6)$~MeV/$c^2$ and $(38\pm 12\pm 2)$~MeV,
respectively, and the statistical significance is more than
$9\sigma$~\cite{omegachic-1}. This structure is in good agreement
with the \Y\ observed in $\eetopipihc$~\cite{y4220_ycz}, and
combined fits assuming the structures at 4.22~GeV/$c^2$ are the
same have been tried by the authors of
Refs.~\cite{combined-fit1,combined-fit2}.

\begin{figure*}[htbp]
\includegraphics[height=5cm]{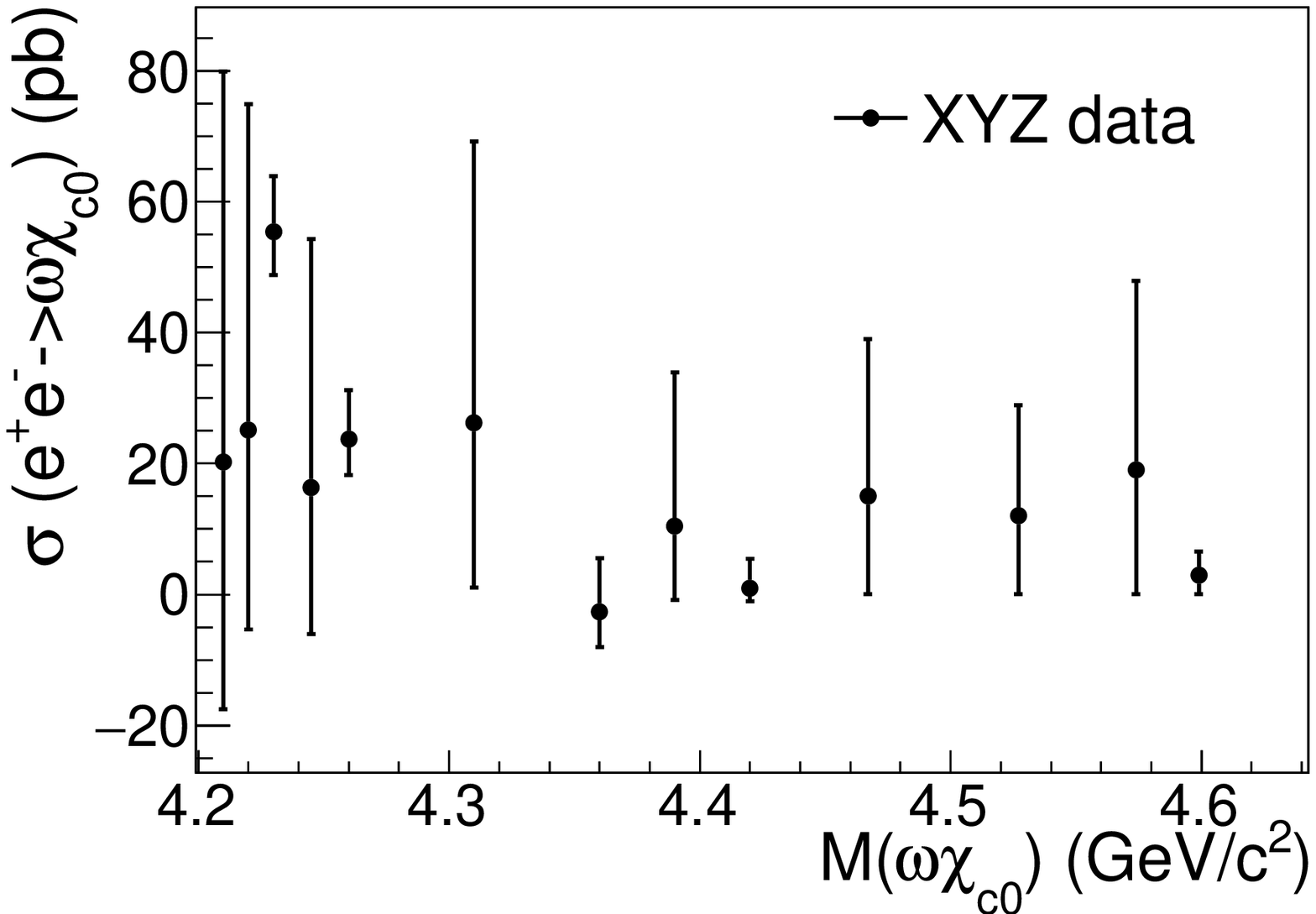}
\includegraphics[height=5cm]{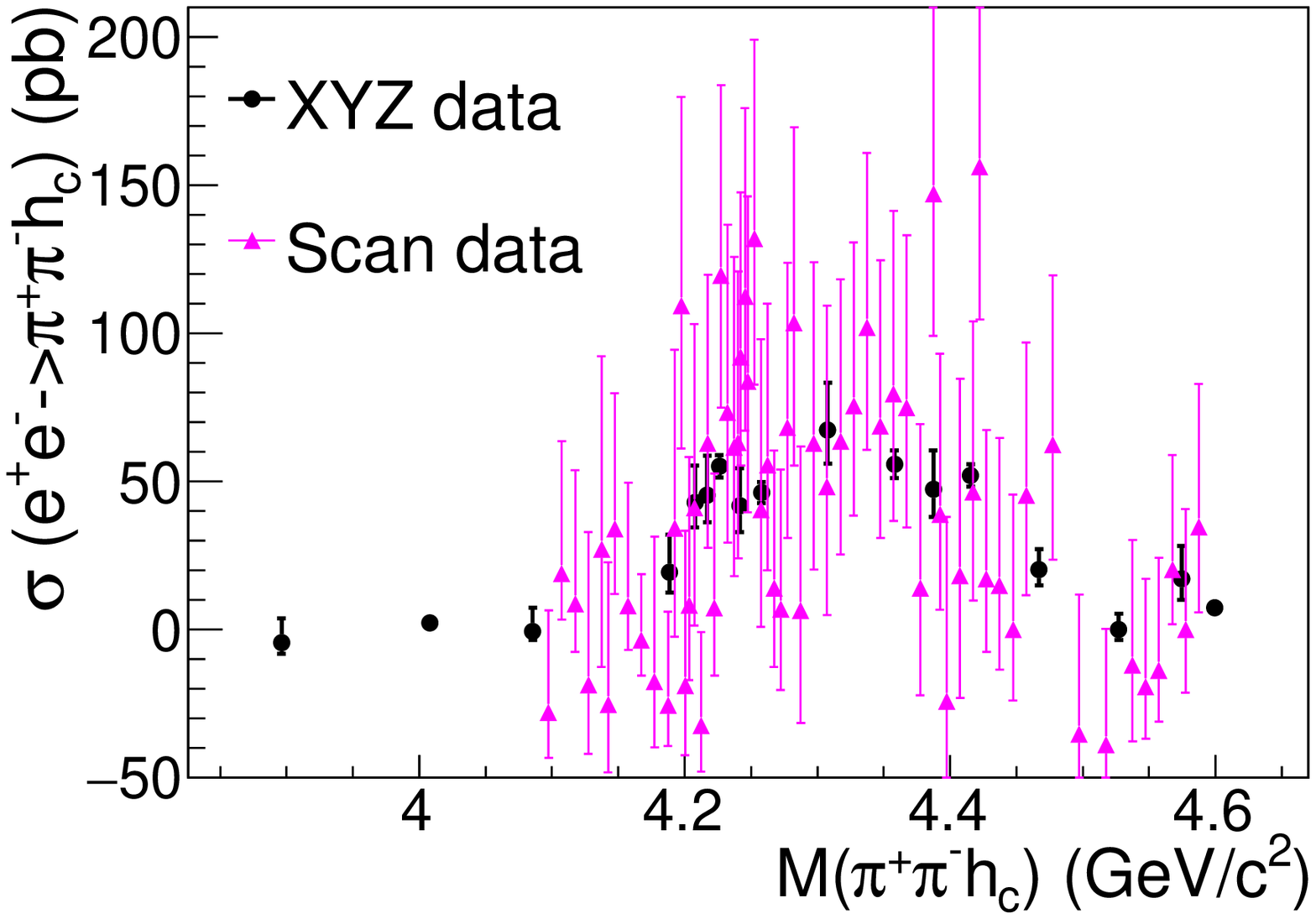}
\includegraphics[height=5cm]{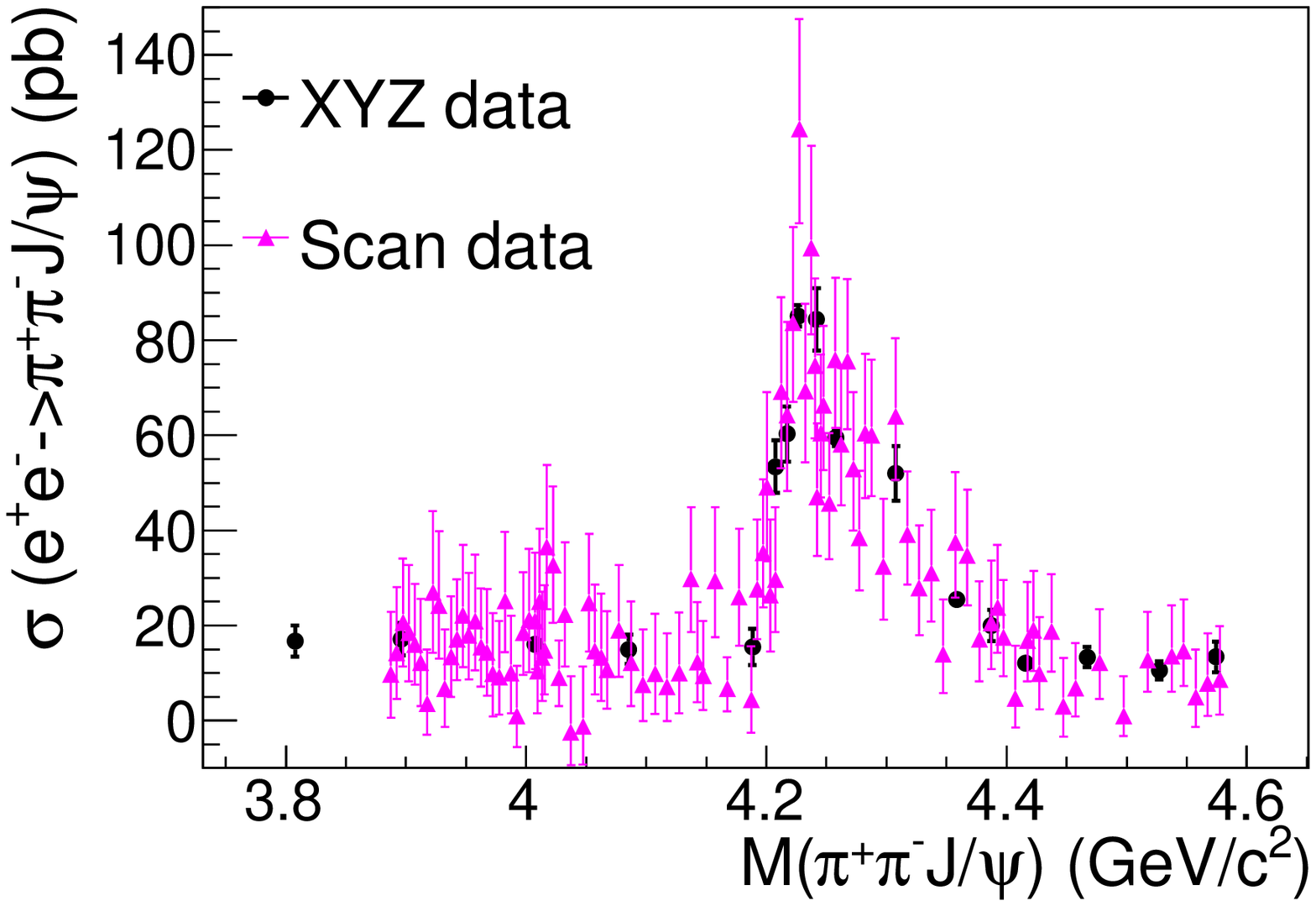}
\includegraphics[height=5cm]{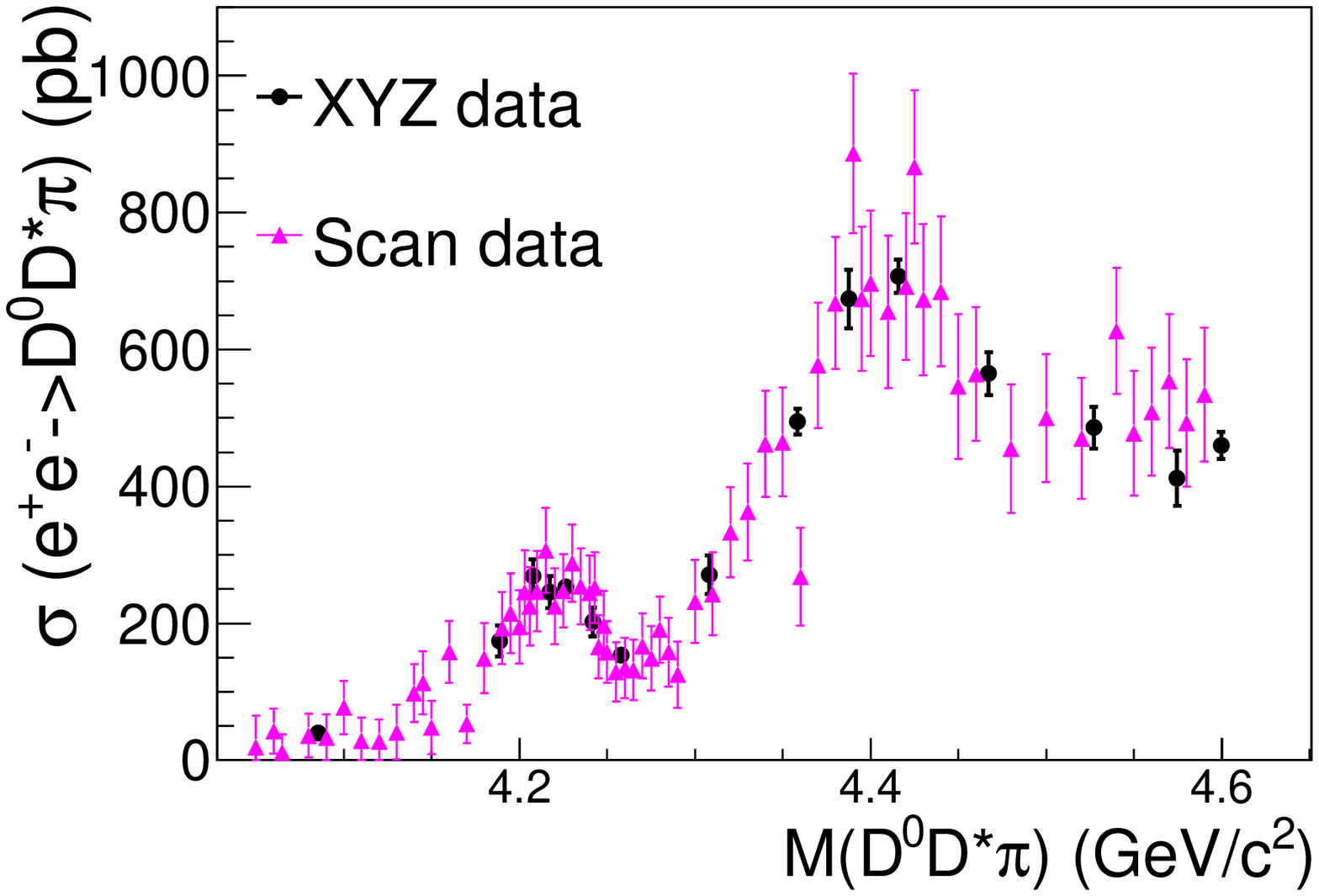}
\caption{The measured cross sections of $e^+e^-\to \omega
\chi_{c0}$, $\pphc$, $\pi^+ \pi^- J/\psi$, and $\ddpi$ by the
BESIII experiment. The dots are from the $XYZ$ data sample and the
triangles are from the $R$-scan data sample. The error bars are
the sum in quadrature of the statistical and uncommon systematic
errors. Here, for each process the correlated systematic uncertainties
(13.3\%,  14.8\%, 5.8\%, and 4.6\% for $\omega
\chi_{c0}$, $\pi^+\pi^-h_c$, $\pi^+\pi^- J/\psi$, and $D^0
D^{*-}\pi^+ + c.c.$, respectively) are
not shown.} \label{omedata}
\end{figure*}

BESIII updated the measurements with higher energy data up to
4.6~GeV included, in both $\eetopipihc$~\cite{pipihc-2} and
$\omega \chi_{c0}$~\cite{omegachic-2} processes. In addition, more
data points are added even at low energy, although with low
integrated luminosity, to further constrain the line shape in
$\eetopipihc$~\cite{pipihc-2} process. While the structure in
$\omega \chi_{c0}$ mode was affected only slightly with the new
measurements at high energies~\cite{omegachic-2}, in the
$\eetopipihc$ mode, the \Y\ was observed with improved
significance together with a new structure, the $Y(4390)$. The
resonant parameters are $M=(4218.4\pm 4.0\pm 0.9)$~MeV/$c^2$ and
$\Gamma=(66.0\pm 9.0\pm 0.4)$~MeV for the \Y, and $M=(4391.6\pm
6.3\pm 1.0)$~MeV/$c^2$ and $\Gamma=(139.5\pm 16.1\pm 0.6)$~MeV for
the $Y(4390)$~\cite{pipihc-2}. The updated cross sections of
$e^+e^- \to \omega \chi_{c0}$ and $\pi^+\pi^- h_c$ are shown in
Fig.~\ref{omedata}, where the measurements at energy points both
with integrated luminosities larger than 40~pb$^{-1}$ (referred to
as `$XYZ$ data sample' hereafter) and with integrated luminosities
smaller than 20~pb$^{-1}$ (referred to as `$R$-scan data sample'
hereafter) are presented.

The process $e^+e^-\to \pi^+\pi^- J/\psi$ at c.m. energies up to
5.0~GeV was first studied by the $BABAR$ experiment, where the
$Y(4260)$ was observed~\cite{y-4260-1}. Belle measured the cross
sections of $e^+e^-\to \pi^+\pi^- J/\psi$ at c.m. energies between
3.8 and 5.0~GeV and reported that $Y(4260)$ alone cannot
describe the line shape satisfactorily~\cite{y4008-1}. Improved
measurements with both $BABAR$~\cite{babar-4008} and
Belle~\cite{y-4260-2-3900-1} full data samples confirmed the existence
of non-$Y(4260)$ component in $e^+e^-\to \pi^+\pi^- J/\psi$ but
the line shape was parametrized with different models. Recently,
BESIII reported a precise measurement of $\eetopipijpsi$ cross
sections at c.m. energies from 3.77 to 4.60~GeV (as shown in
Fig.~\ref{omedata}) using data samples with an integrated
luminosity of 9~fb$^{-1}$~\cite{pipijpsi}. While the nature of the
events at around 4~GeV is still ambiguous, the dominant resonant
structure, the so called $Y(4260)$, was found to have a mass of
$(4222.0\pm 3.1\pm 1.4)$~MeV/$c^2$ and a width of $(44.1\pm 4.3\pm
2.0)$~MeV, in good agreement with the \Y\ observed in
$\eetopipihc$~\cite{pipihc-2}. In addition, a new resonance with a
mass of around 4.32~GeV/$c^2$ is needed to describe the high
precision data.

BESIII also reported a measurement of the $\eetoddpi$ cross
sections at c.m. energies from 4.05 to 4.60~GeV with the same data
samples~\cite{ddpi-bes}, which is a significant improvement over
the previous measurement at Belle~\cite{ddpi-belle}. Two resonant
structures in good agreement with the \Y\ and $Y(4390)$ observed
in $\pphc$~\cite{pipihc-2} are identified over a smoothly
increasing non-resonant term which can be parametrized with a
three-body phase space amplitude. The cross sections of $\ddpi$
are also shown in Fig.~\ref{omedata}.

An obvious feature in the above four channels from the BESIII
measurements is that there is a common structure at around
4.22~GeV/$c^2$, i.e., the \Y. As such a state is not observed in
other open charm final states~\cite{galina}, these four final
states are probably the dominant decay modes of the \Y. By
applying constraints to the resonant parameters in a simultaneous
fit to the cross sections of these four processes, we may obtain
the best knowledge on the \Y, including resonant parameters (mass,
width, coupling to lepton pair, and decay branching fractions),
and thus a better understanding of its nature~\cite{xyz-review},
especially whether it is an exotic state, such as a tetraquark
state in the diquark-antidiquark model~\cite{1412.7196}, a vector
molecular state of $D\bar{D_1}(2420)$~\cite{1310.2190}, a mixture
of two hadrocharmonium states~\cite{voloshin}, an
$\omega\chi_{cJ}$ molecule~\cite{zhenghq,yuancz}, or a
charmonium-hybrid state~\cite{ccg_lqcd}.

\section {The data and the fit formalism}

We use the measured cross sections of $e^+e^- \to \omega
\chi_{c0}$, $\pi^+\pi^-h_c$,  $\ppjpsi$, and $\ddpi$
processes~\cite{omegachic-2, pipihc-2, pipijpsi, ddpi-bes} by
BESIII experiment only to measure the parameters of the resonances
presented. The data are shown in Fig.~\ref{omedata}, where the
dots with error bars are from the $XYZ$ data sample and the
triangles with error bars are from the $R$-scan data sample. Here,
the error bars are the sum in quadrature of the statistical and uncorrelated
systematic errors, and the correlated systematic uncertainties common to
the energy points in each process are removed since they have no
effect on the fitted resonant parameters.

We parametrize the cross section with the coherent sum of a few
amplitudes, either resonance represented by a Breit-Wigner (BW)
function or non-resonant production term parametrized with a phase
space term. The BW function used in this article is
$$
BW(\sqrt{s})=\frac{\sqrt{12\pi \Gamma_{e^+e^-} \BR_f \Gamma}}
{s-M^2+iM\Gamma}\sqrt{\frac{PS_n(\sqrt{s})}{PS_n(M)}},
$$
where $M$ is the mass of the resonance; $\Gamma$ and
$\Gamma_{e^+e^-}$ are the total width and partial width to
$e^+e^-$, respectively; $\BR_f$ is the branching fraction of the
resonance decays into final state $f$; and $PS_n$ is the $n-$body
decay phase space factor which increases smoothly from the mass threshold with the
$\sqrt{s}$~\cite{pdg}.

In fitting to the data shown in Fig.~\ref{omedata}, we assume the
observed structures at around 4.22~GeV/$c^2$ in all reactions and
structures at around 4.39~GeV/$c^2$ in $\eetopipihc$ and $\ddpi$
are due to the same resonant states, i.e., we assume the cross
sections are due to the \Y\ only for $\omega \chi_{c0}$, the \Y\
and $Y(4390)$ for $\pi^+\pi^- h_c$, the $Y(4008)$, \Y\, and
$Y(4320)$ for $\pi^+ \pi^- J/\psi$, and the \Y\ and $Y(4390)$ for
$\ddpi$, that is,
\begin{eqnarray}
\sigma_{\omega\chi_{c0}}(\sqrt{s}) &=& |BW_1(\sqrt{s})|^2,
                        \label{omega-eq} \\
  \sigma_{\pi^+\pi^- h_c}(\sqrt{s}) &=&
  |BW_1(\sqrt{s})+BW_3(\sqrt{s})\cdot e^{i\phi_1}|^2,
                        \label{pipihc-eq} \\
  \sigma_{\ppjpsi}(\sqrt{s}) &=&
  |BW_0(\sqrt{s})+BW_1(\sqrt{s})\cdot e^{i\phi_2}+ BW_2(\sqrt{s})\cdot
  e^{i\phi_3}|^2,
                        \label{pipijpsi-eq} \\
  \sigma_{\ddpi}(\sqrt{s}) &=& |\sqrt{PS_3(\sqrt{s})}+BW_1(\sqrt{s})\cdot
  e^{i\phi_4}+BW_3(\sqrt{s})\cdot e^{i\phi_5}|^2,
                        \label{ddpi-eq}
\end{eqnarray}
where $BW_0$, $BW_1$, $BW_2$, and $BW_3$ represent the $Y(4008)$,
\Y, $Y(4320)$, and $Y(4390)$, respectively, and $\phi$ is the
relative phase between the amplitudes.

We do a combined fit using a least squares method with {\sc
minuit} package in the CERN Program Library~\cite{minuit}. The
$\chi^2$ function is constructed as
$$
\chi^2 =\sum^{4}_{j=1} \sum^{n}_{i=1} \frac{(\sigma^{data}_{ij} -
\sigma^{fit}_{ij})^2}{\delta_{ij}^2},
$$
where $\sigma^{data}_{ij}$ and $\sigma^{fit}_{ij}$ are the
measured and fitted cross sections of the $i$th energy point in
the $j$th mode, $\delta_{ij}$ is the corresponding total error
with common systematic errors removed. The sum is performed over
all the measured cross section points from the above mentioned
four modes. The $\chi^2$ is minimized to obtain the best
estimation of the resonant parameters.

\section{Fit results}

We fit BESIII data on $e^+e^- \to \omega \chi_{c0}$, $\pi^+\pi^-
h_c$, $\pi^+ \pi^- J/\psi$, and $\ddpi$ cross sections
simultaneously. Two solutions, four solutions, and four solutions
with the same minimum values of $\chi^2$ are found with the two,
three, and three amplitudes interfering with each other for
$e^+e^- \to  \pi^+\pi^- h_c$, $\pi^+ \pi^- J/\psi$, and $\ddpi$,
respectively. The masses and the widths of the resonances are
identical but the partial widths to $e^+e^-$ and relative phases
are different in different solutions for each process. There are
no multiple solutions for $e^+e^- \to \omega \chi_{c0}$ since only
one amplitude is used.

\begin{figure*}[htbp]
\hbox{
 \psfig{file=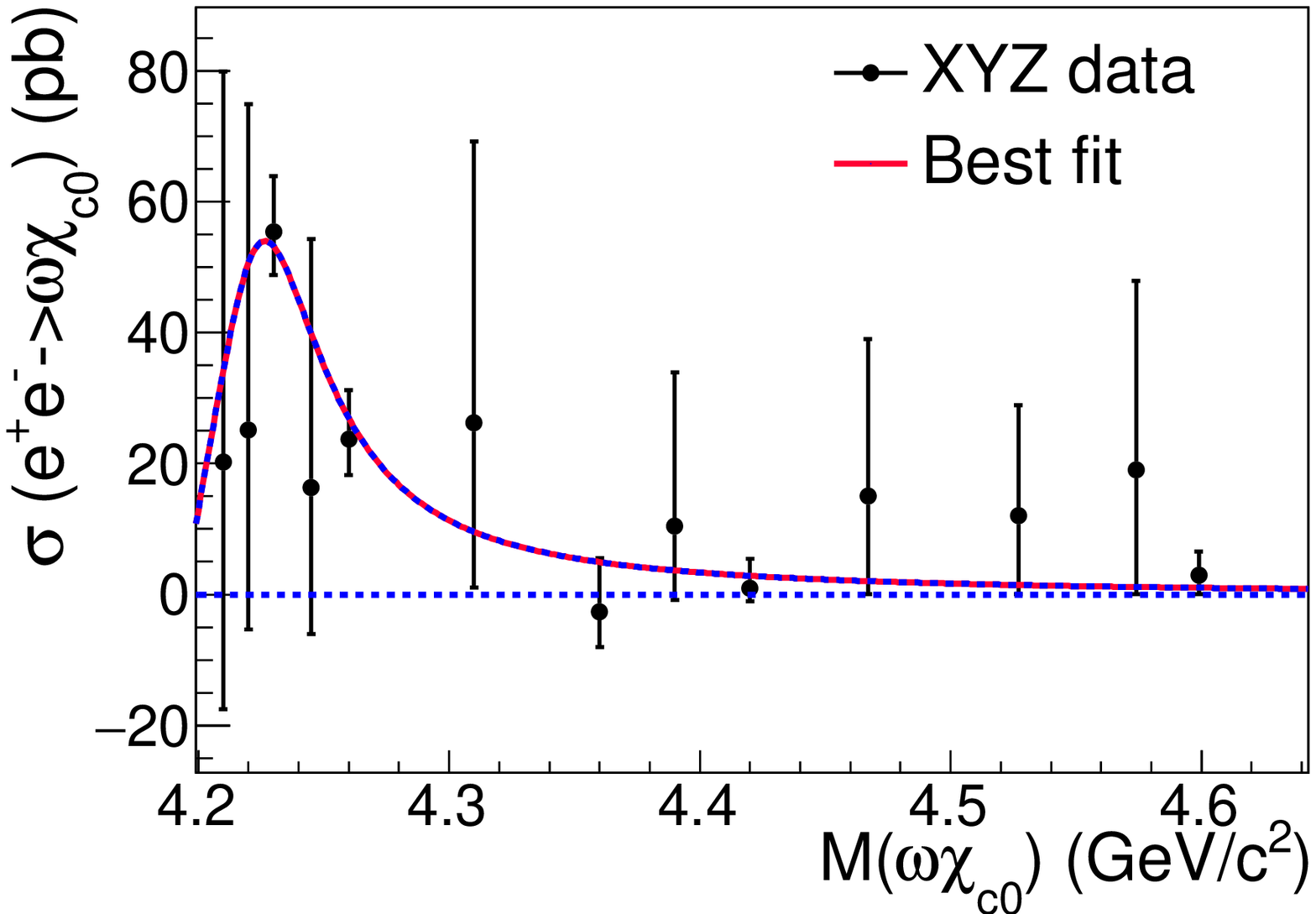,width=4cm, angle=0}
 }
\hbox{
 \psfig{file=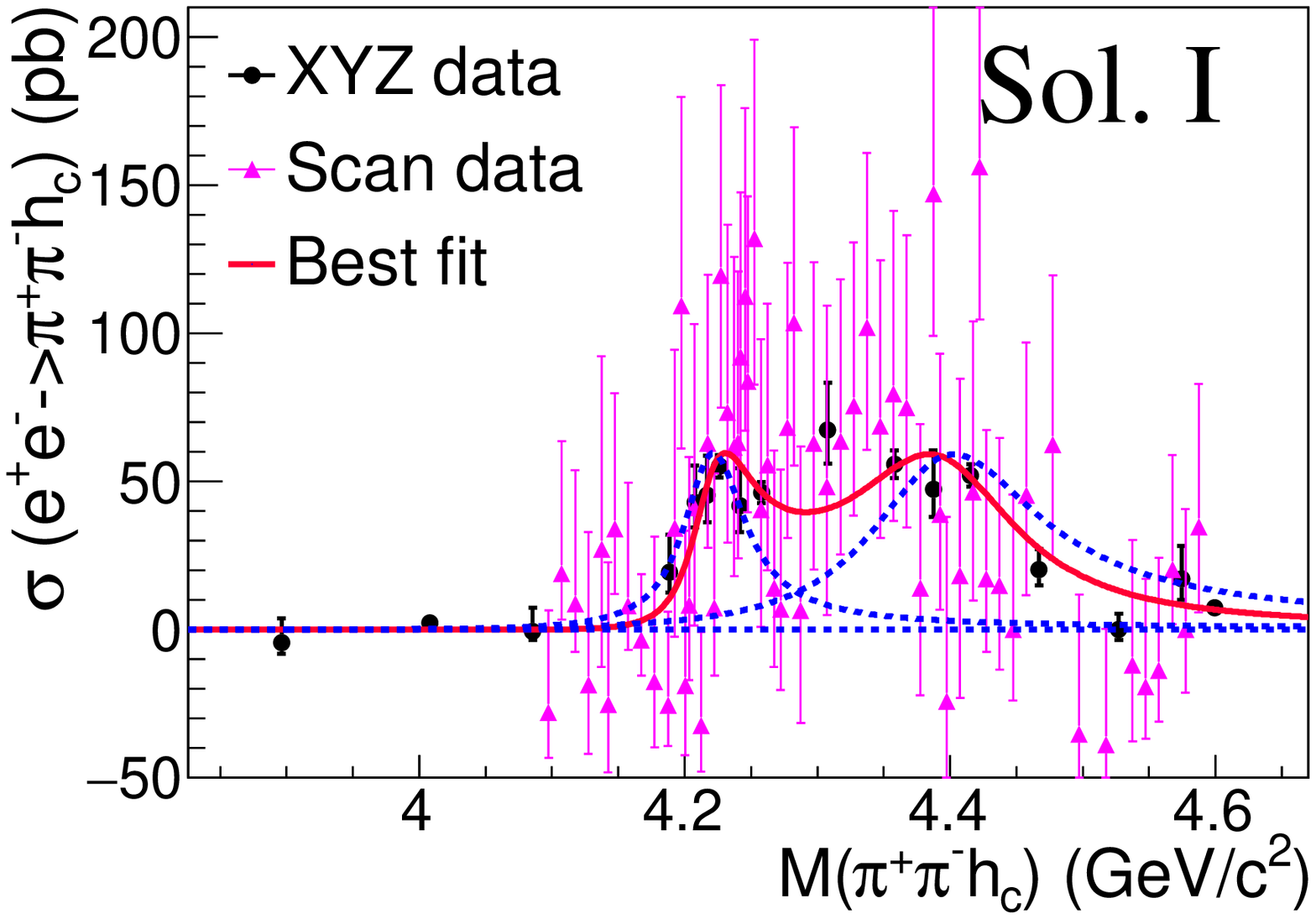,width=4cm, angle=0}
 \psfig{file=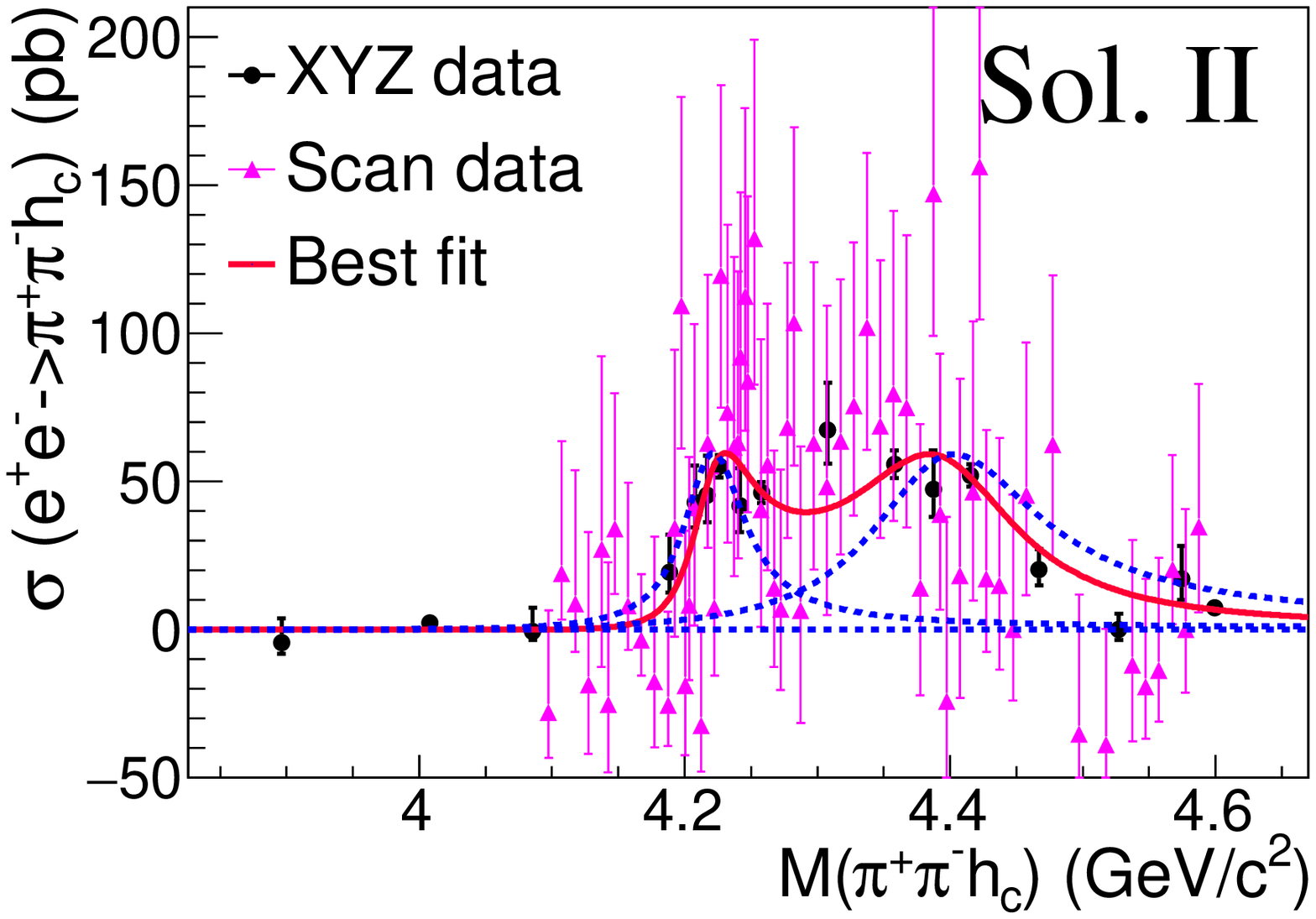,width=4cm, angle=0}
 }
\hbox{
 \psfig{file=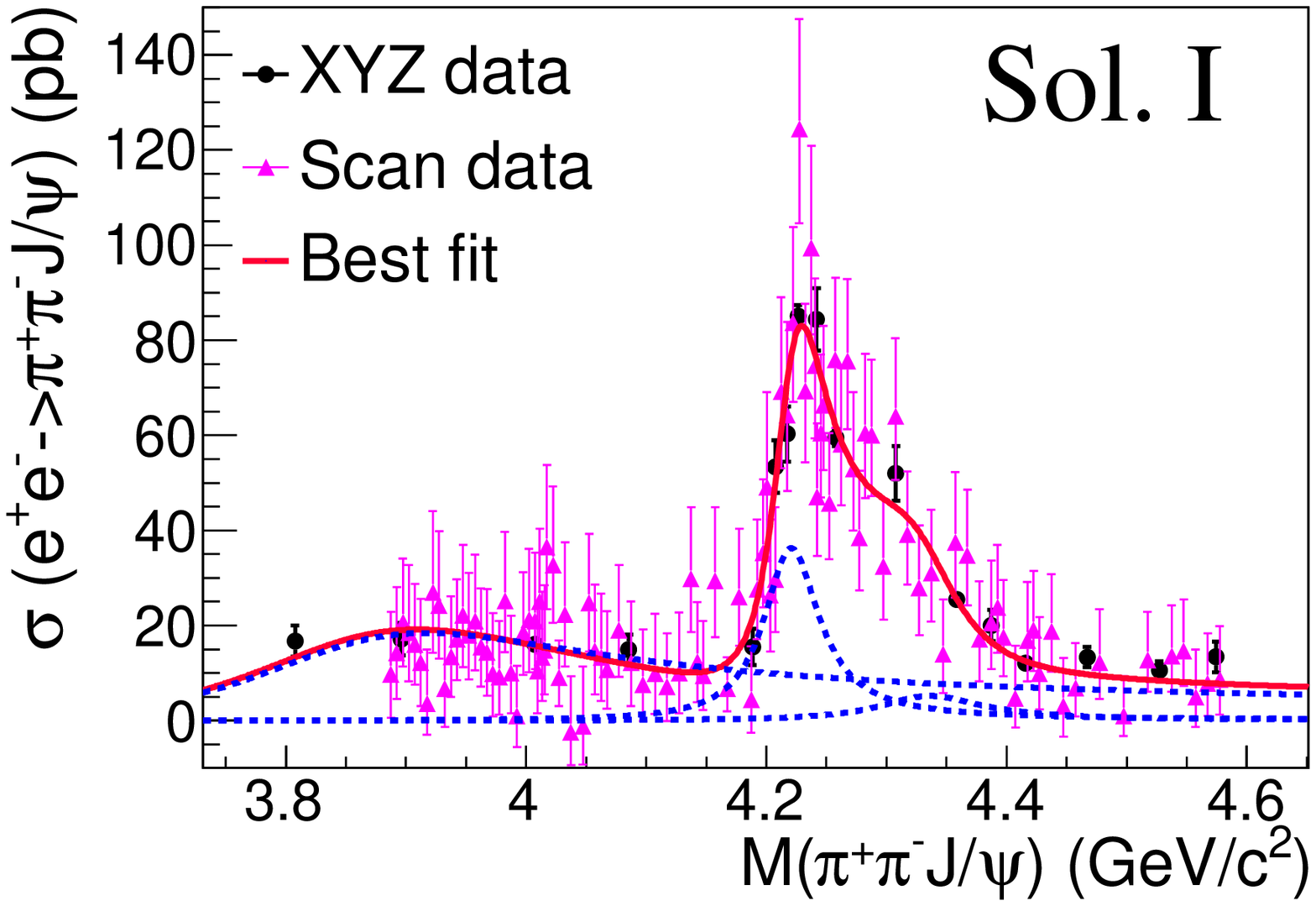,width=4cm, angle=0}
 \psfig{file=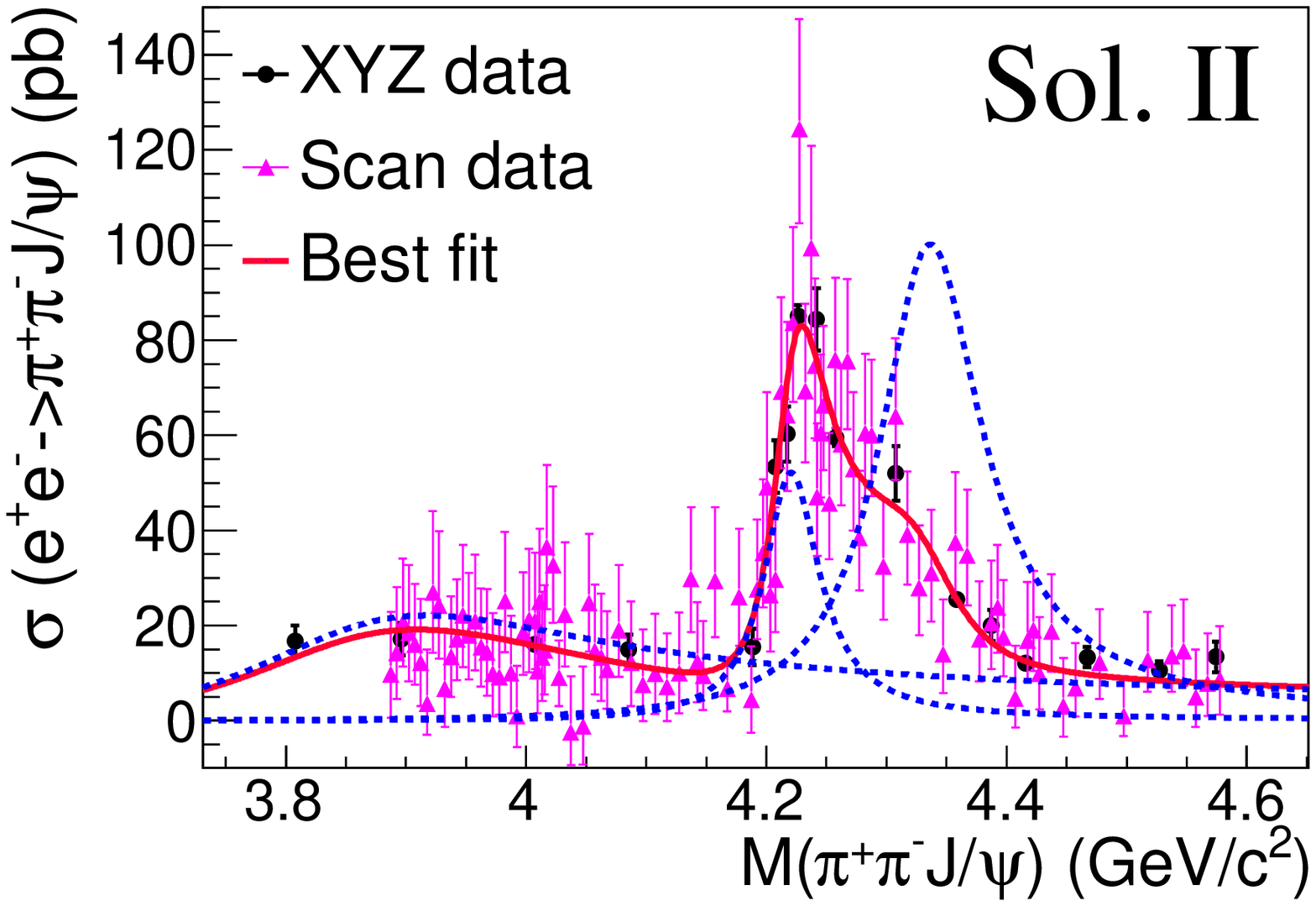,width=4cm, angle=0}
 \psfig{file=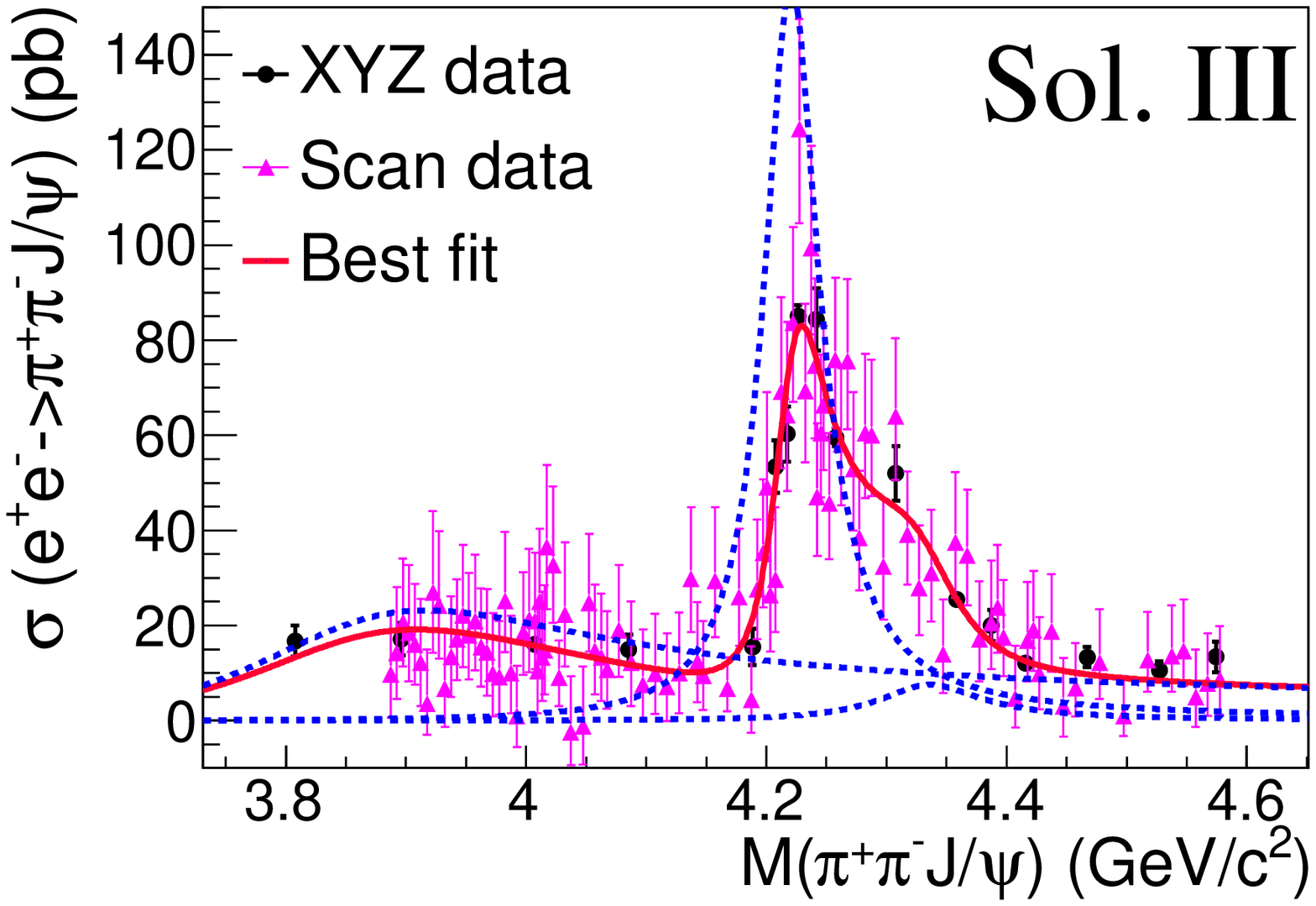,width=4cm, angle=0}
 \psfig{file=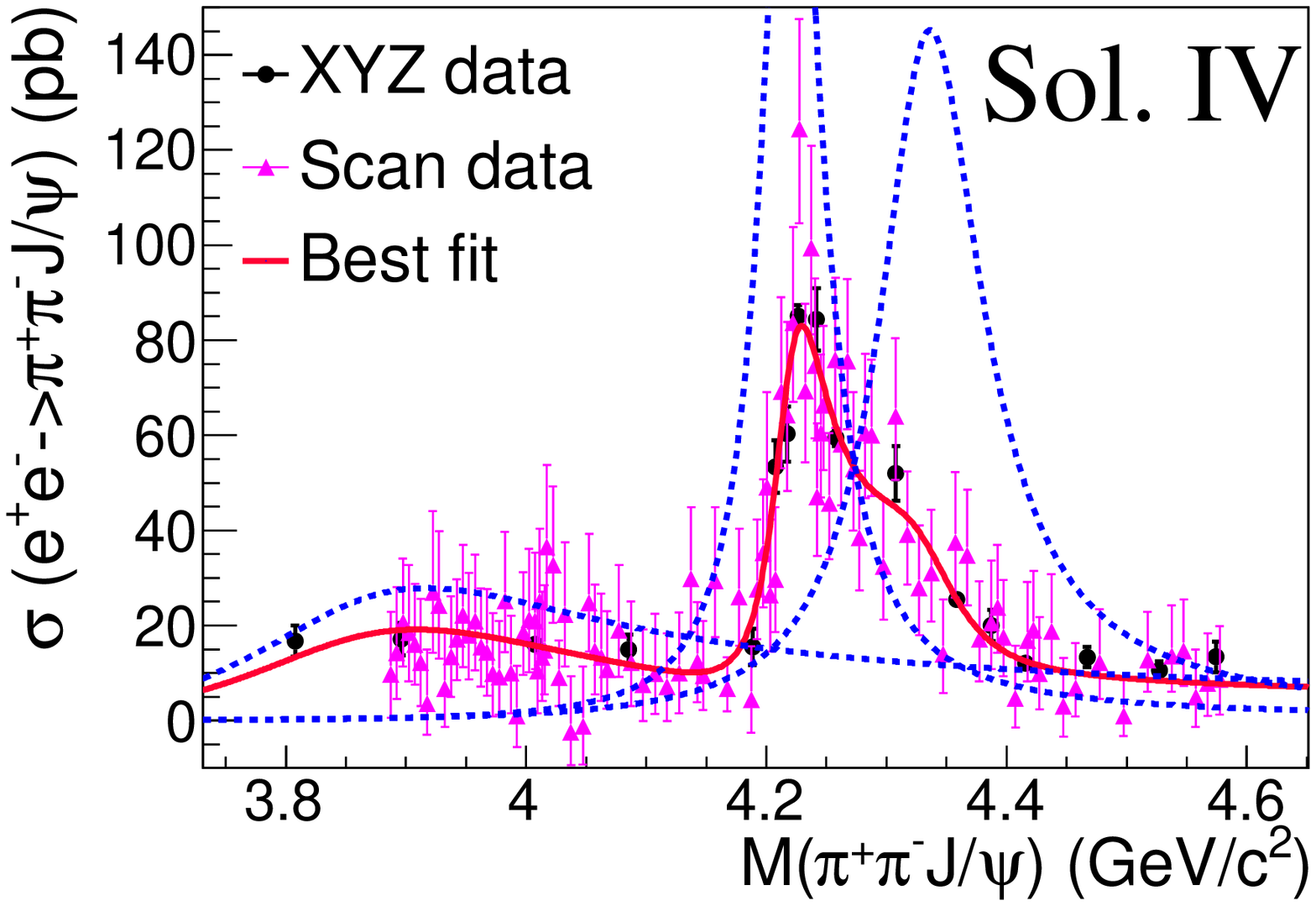,width=4cm, angle=0}
 }
\hbox{
 \psfig{file=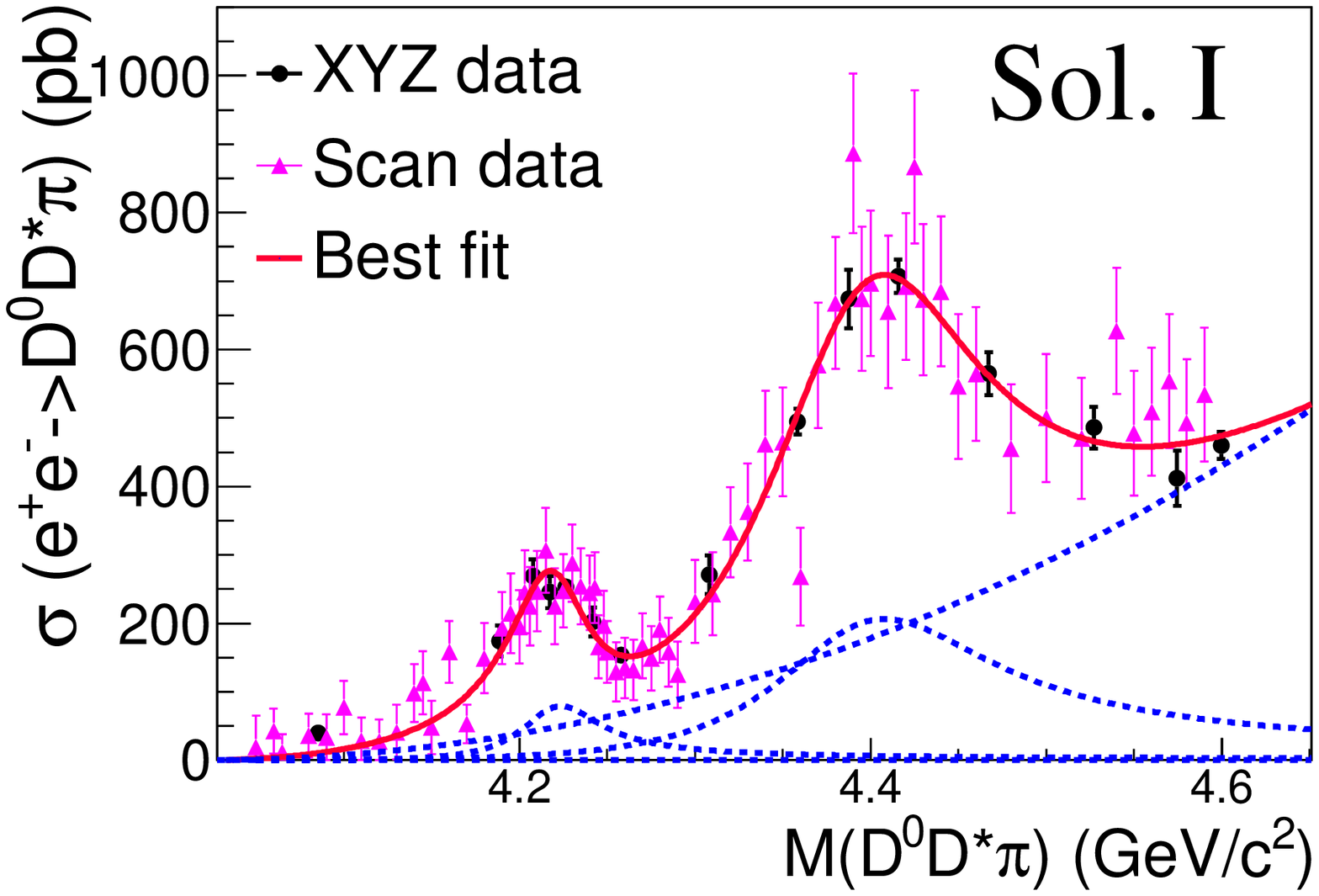,width=4cm, angle=0}
 \psfig{file=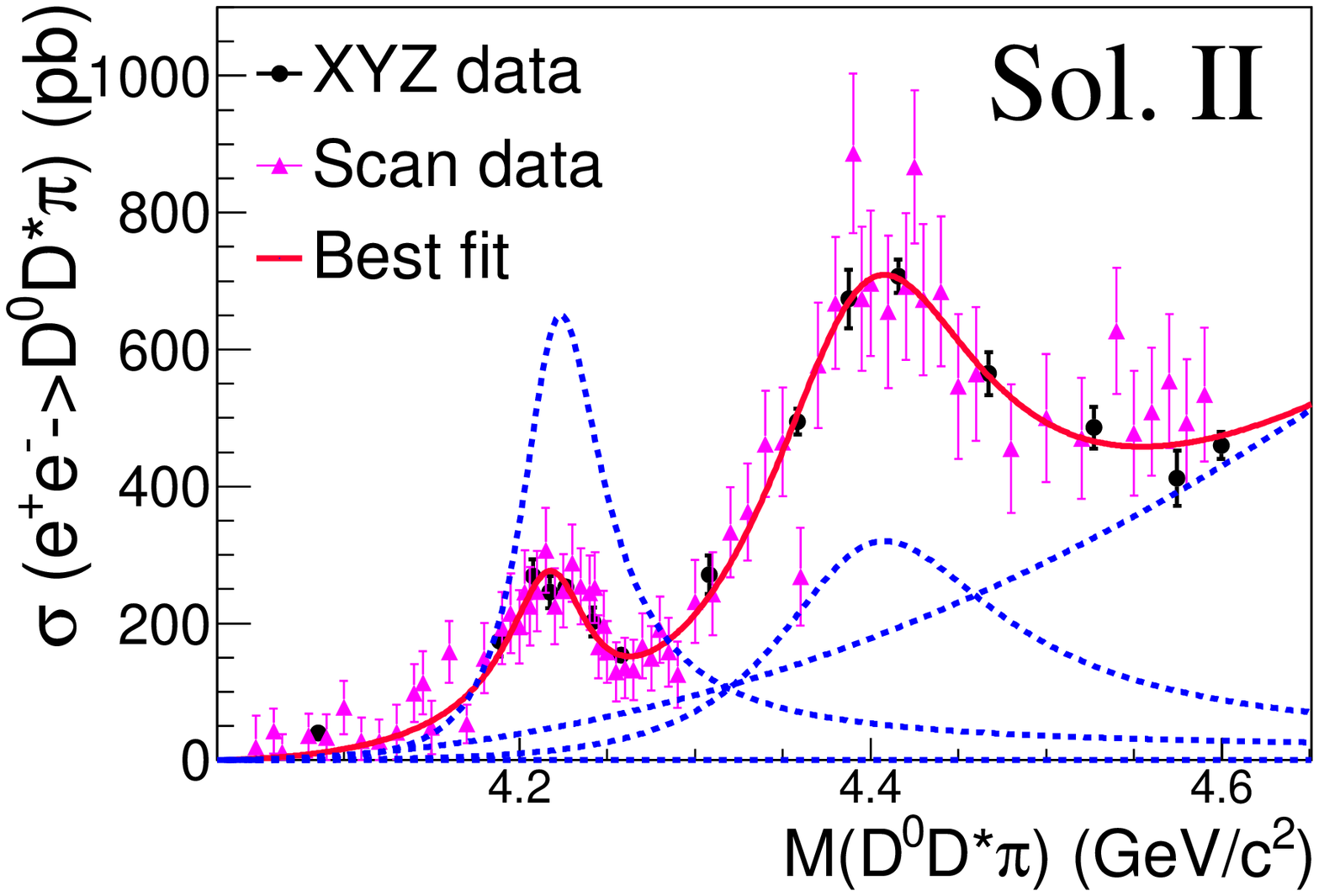,width=4cm, angle=0}
 \psfig{file=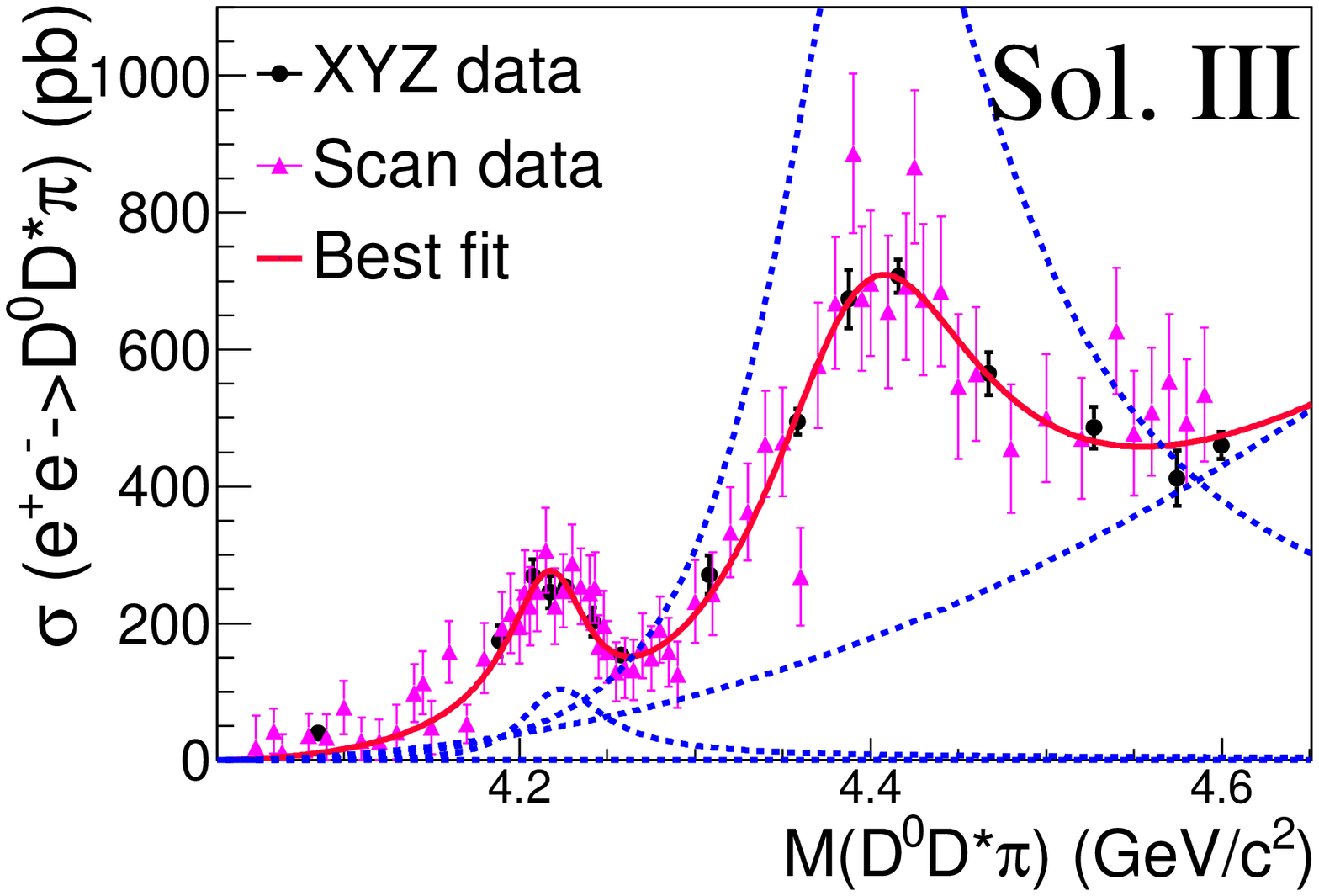,width=4cm, angle=0}
 \psfig{file=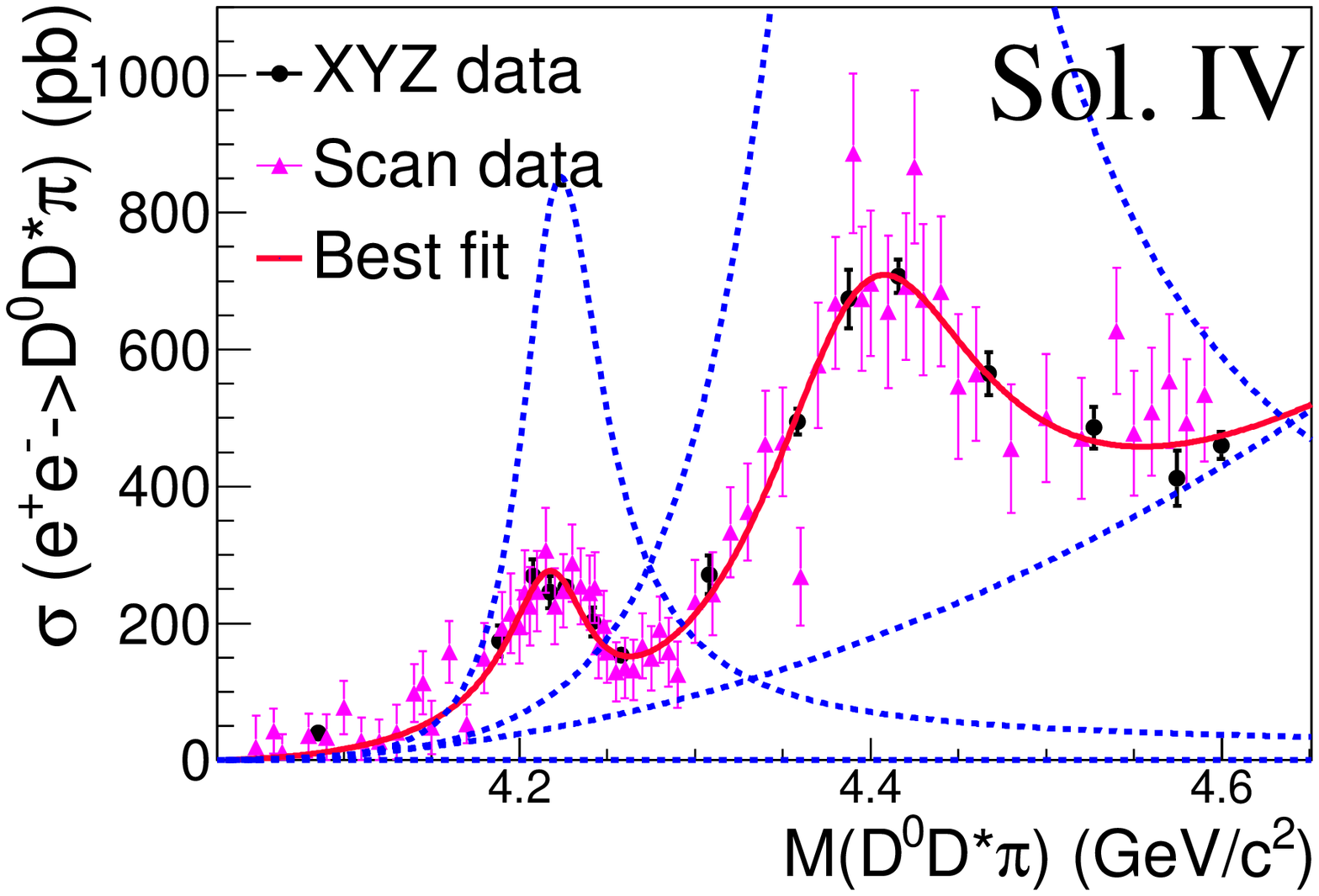,width=4cm, angle=0}
 }
\caption{ The results of the combined fit to $\EE\to
\omega\chi_{c0}$, $\pphc$, $\pi^+ \pi^- J/\psi$, and $\ddpi$ (from
the top to the bottom row). The dots and the triangles with errors
bars are data as described in Fig.~\ref{omedata} caption. The
solid curves are the projections from the best fit. The dashed
curves show the fitted resonance components from different
solutions indicated in the top right corner in each plot. The
numerical results of all the solutions are presented in
Table~\ref{ddpi-table}.} \label{result-fit}
\end{figure*}

Figure~\ref{result-fit} shows the fit results with a
goodness-of-the-fit of $\chi^2 / ndf$ = 241/273=0.9, where the
solid curves show the projections from the best fit, the dashed
curves show the fitted resonance components from different
solutions, and the corresponding mass, width, and the product of
the branching fraction to specific mode and the $e^+e^-$  partial
width for each resonance are listed in Table~\ref{ddpi-table}.

\begin{table*}[htbp]
\caption{The resonant parameters from the combined fit to $\EE\to
\omega\chi_{c0}$, $\pphc$, $\pi^+ \pi^- J/\psi$, and $\ddpi$. Here
$M$, $\Gamma$, and $(\BR_{i}\times \Gamma_{e^+e^-})_j$ are the
mass (in MeV/$c^2$), total width (in MeV), and the product of the
branching fraction to specific final state and the $e^+e^-$
partial width (in eV), respectively, where $i$ presents a final
state and $j$ indicates a resonance added in the fit in different
processes. The fitted mass and width for each resonance are shown
in the upper table separately. All the errors are statistical from
fit only. $\phi$ is the relative phase (in rad).}
{\begin{tabular}{@{}ccccc@{}}\toprule
                        &~~~~~~~~$Y(4008)$~~~~~~~&~~~~~~~\Y~~~~~~~~~&~~~~~~~$Y(4320)$~~~~~~~~~&~~~~~~~$Y(4390)$~~~~~~~~~\\
\colrule
\hphantom{0}$M$     &   $3846.3 \pm 45.5$   &    $4219.6 \pm 3.3$   &   $4333.2 \pm 19.9$   &   $4391.5 \pm  6.3$\\
\hphantom{0}$\Gamma$&    $345.6 \pm 58.2$   &      $56.0 \pm 3.6$   &    $104.3 \pm 44.9$   &    $153.2 \pm 11.4$\\
\botrule\\
\end{tabular}
\vspace{-0.6cm}
\begin{center}
\begin{tabular}{@{}ccccc@{}}\toprule
&~~Solution I~~&~~Solution II~~&~~Solution III~~&~~Solution IV~~\\
\colrule

%\multicolumn{5}{c}{$\eetoomegachic$} \\
%\colrule
$(\BR_{\omega \chi_{c0}}\times \Gamma_{e^+e^-})_{Y(4220)}$\hphantom{00}    & \hphantom{0}$3.4\pm 0.4$ \\
\colrule

%\multicolumn{5}{c}{$\eetopipihc$} \\
%\colrule
$(\BR_{\pi^+\pi^-h_c}\times \Gamma_{e^+e^-})_{Y(4220)}$\hphantom{00}    & \hphantom{0}$4.0\pm 1.1$ & \hphantom{0}$4.0\pm1.1$ \\

$(\BR_{\pi^+\pi^-h_c}\times \Gamma_{e^+e^-})_{Y(4390)}$\hphantom{00}    & \hphantom{0}$11.7\pm 2.4$ & \hphantom{0}$11.7\pm2.5$ \\
$\phi_1$\hphantom{00}                                           & \hphantom{0}$3.1\pm 0.4$ & \hphantom{0}$-3.2\pm0.4$ \\
\colrule

%\multicolumn{5}{c}{$\eetopipijpsi$} \\
%\colrule

$(\BR_{\pi^+\pi^-J/\psi}\times \Gamma_{e^+e^-})_{Y(4008)}$\hphantom{00}    & \hphantom{0}$5.5\pm 0.3$ & \hphantom{0}$6.6\pm0.7$    & \hphantom{0}$6.9\pm 0.7$ & \hphantom{0}$8.3\pm0.7$\\

$(\BR_{\pi^+\pi^-J/\psi}\times \Gamma_{e^+e^-})_{Y(4220)}$\hphantom{00}    & \hphantom{0}$2.5\pm 0.2$ & \hphantom{0}$3.5\pm0.7$    & \hphantom{0}$10.5\pm1.1$ & \hphantom{0}$15.1\pm1.3$\\
$\phi_2$\hphantom{00}                                               & \hphantom{0}$0.1\pm 0.1$ & \hphantom{0}$0.8\pm0.3$    & \hphantom{0}$-1.8\pm 0.2$ & \hphantom{0}$-1.0\pm0.1$\\

$(\BR_{\pi^+\pi^-J/\psi}\times \Gamma_{e^+e^-})_{Y(4320)}$\hphantom{00}    & \hphantom{0}$0.7\pm 0.2$ & \hphantom{0}$13.3\pm3.8$  & \hphantom{0}$1.0\pm 0.5$ & \hphantom{0}$19.4\pm3.2$\\
$\phi_3$\hphantom{00}                                               & \hphantom{0}$2.2\pm 0.2$ & \hphantom{0}$-2.0\pm0.2$  & \hphantom{0}$1.4\pm 0.6$ & \hphantom{0}$-2.7\pm0.1$\\
\colrule

%\multicolumn{5}{c}{$\eetoddpi$} \\
%\colrule

$(\BR_{\ddpi}\times \Gamma_{e^+e^-})_{Y(4220)}$\hphantom{00}    & \hphantom{0}$5.3\pm 0.6$ & \hphantom{0}$43.3\pm3.2$    & \hphantom{0}$6.9\pm 0.8$ & \hphantom{0}$56.7\pm4.2$\\
$\phi_4$\hphantom{00}                                       & \hphantom{0}$2.2\pm 0.1$ & \hphantom{0}$-2.2\pm0.1$    & \hphantom{0}$-2.7\pm 0.1$ & \hphantom{0}$-0.8\pm0.1$\\

$(\BR_{\ddpi}\times \Gamma_{e^+e^-})_{Y(4390)}$\hphantom{00}    & \hphantom{0}$39.7\pm 4.3$ & \hphantom{0}$61.6\pm6.6$  & \hphantom{0}$265.5\pm 16.6$ & \hphantom{0}$412.0\pm26.0$\\
$\phi_5$\hphantom{00}                                       & \hphantom{0}$1.9\pm 0.1$ & \hphantom{0}$1.5\pm0.2$    & \hphantom{0}$4.7\pm 0.1$ & \hphantom{0}$4.2\pm0.1$\\
\botrule
\end{tabular}
\end{center}
\label{ddpi-table}}
\end{table*}

From the fit we obtain $M=(4219.6 \pm 3.3)$~MeV/$c^2$ and
$\Gamma=(56.0\pm 3.6)$~MeV for the \Y\, where the errors are
combined statistical and uncorrelated systematic uncertainties. We can find that the resonant parameters are
significantly different from those of the $Y(4260)$ determined
from low statistics experiments $BABAR$~\cite{babar-4008} and
Belle~\cite{y-4260-2-3900-1}, although they are obviously the same
resonant structure.

\section{Systematic errors}

The systematic uncertainties in the resonant parameters in the
combined fit to the cross sections of $e^+e^-\to \omega
\chi_{c0}$, $\pi^+\pi^-h_c$, $\pi^+\pi^- J/\psi$, and $D^0
D^{*-}\pi^+ + c.c.$ are mainly from the absolute c.m. energy
measurement, the c.m. energy spread, cross section measurements, parametrization of the
resonances and background shape.

The systematic uncertainties on the  mass and width from the absolute c.m. energy measurement and
the c.m. energy spread are taken from the original BESIII
publications~\cite{omegachic-2, pipihc-2, pipijpsi, ddpi-bes}, where we take the
largest values conservatively when they are different in different modes. The uncertainty from the cross section measurement in each mode
is divided into two categories, correlated and uncorrelated systematic uncertainties.
For the uncorrelated systematic uncertainties, they are added with the statistical errors
in quadrature, as shown in Figs.~\ref{omedata} and ~\ref{result-fit}, i.e., the fit errors
have covered uncorrelated systematic uncertainties in the cross sections.
The correlated uncertainty from the cross section measurement in each mode is
common for all data points~\cite{omegachic-2, pipihc-2, pipijpsi,
ddpi-bes}, which only affects the $\BR \times \Gamma_{e^+e^-}$
measurement and is 13.3\%,  14.8\%, 5.8\%, and 4.6\% for $\omega
\chi_{c0}$, $\pi^+\pi^-h_c$, $\pi^+\pi^- J/\psi$, and $D^0
D^{*-}\pi^+ + c.c.$, respectively.

Instead of using a constant total width, we assume a mass
dependent width to estimate the uncertainty due to signal
parametrization. To model the $\pi^+\pi^- J/\psi$ cross section
near 4~GeV, an exponential function as used in
Ref.~\cite{pipijpsi} is taken instead of using the $Y(4008)$
resonance. We consider the systematic bias introduced by possible
additional resonances in the processes under study. The fit
scenarios include adding an additional phase space term for
$\omega \chi_{c0}$; using three resonances, the \Y, $Y(4320)$ and
$Y(4390)$, to fit $\pi^+\pi^-h_c$, $D^0 D^{*-}\pi^+ + c.c.$, or
both of them. The shifts of the masses and widths are taken as
systematic uncertainties.

The overall systematic uncertainties are obtained by adding all
the sources of systematic uncertainties in quadrature assuming
they are independent, which are 16.7~MeV/$c^2$ and 31.6~MeV for
the mass and width of the $Y(4008)$, respectively; 5.1~MeV/$c^2$
and 6.9~MeV for the mass and width of the \Y, respectively;
20.9~MeV/$c^2$ and 23.1~MeV for the mass and width of the
$Y(4320)$, respectively; and 20.8~MeV/$c^2$ and 16.4~MeV for the
mass and width of the $Y(4390)$, respectively.

\section{Summary and discussions}

From a combined fit to the $e^+e^- \to \omega \chi_{c0}$, $\pphc$,
$\pi^+ \pi^- J/\psi$, and $\ddpi$ cross sections measured by
BESIII, we determine the mass of the \Y\ as $(4219.6 \pm 3.3 \pm
5.1)$~MeV/$c^2$ and the width of $(56.0 \pm 3.6 \pm 6.9)$~MeV, and
the relative production rates in these four decay modes.

The leptonic decay width for a vector state is an important
quantity for discriminating various theoretical interpretations
of its nature. The magnitude of the leptonic decay width
determines how the strong decay widths sum up to the total width.
Smaller leptonic decay width means that the strong decay widths
will be relatively enhanced and vice versa. As the \Y\ is the
dominant component in $\EE\to \pi^+\pi^-J/\psi$, we assume the
theoretical interpretations of $Y(4260)$ may apply also for the \Y.

The recent estimate of Lattice QCD (LQCD) for the leptonic decay
width of the \Y\ is about 40~eV~\cite{chenying} as a feature of
the hybrid scenario; the predicted upper limit of the \Y\ leptonic
decay width is about 500~eV if the \Y\ is a hadronic molecule
dominated by $D\bar{D}_1(2420)$~\cite{mole}; the leptonic
decay width is only about 23~eV for the $\omega\chi_{c0}$ molecule
interpretation~\cite{dai} where no contributions from the open
charm decay channel are included in the analysis.

By considering the isospin symmetric modes of the measured
channels, we can estimate the lower limit on the leptonic partial
width of the \Y\ decays. For an isospin-zero charmoniumlike state,
we expect
\begin{eqnarray*}
  \BR_{\pi\pi h_c} &=& \frac{3}{2}\times \BR_{\pi^+\pi^- h_c}, \\
  \BR_{\pi\pi J/\psi} &=& \frac{3}{2}\times \BR_{\pi^+\pi^- J/\psi},  \\
  \BR_{D\bar{D^*}\pi} &=& 3\times \BR_{\ddpi}.
\end{eqnarray*}
So we have
\begin{eqnarray*}\label{gee}
 \Gamma_{e^+e^-} &=& \sum_i \BR_{i}\times \Gamma_{e^+e^-} \\
                 &=& \BR_{\omega\chi_{c0}}\times \Gamma_{e^+e^-}
                 + \BR_{\pi\pi h_c}\times \Gamma_{e^+e^-}
                 + \BR_{\pi\pi J/\psi}\times \Gamma_{e^+e^-}
                 + \BR_{D\bar{D^*}\pi}\times \Gamma_{e^+e^-}
                 + ...
\end{eqnarray*}

By inserting the numbers from Table~\ref{ddpi-table}, considering the
solutions with the smallest $\BR\times
\Gamma_{e^+e^-}$, we obtain
\begin{eqnarray*}\label{gee0}
 \Gamma_{e^+e^-} &=& (3.4\pm 0.4 \pm 1.8)
                 + \frac{3}{2}\times (4.0\pm 1.1 \pm 3.2)
                 + \frac{3}{2}\times (2.5\pm 0.2 \pm 0.8) \\
                 && +\, 3\times (5.3\pm 0.6 \pm 1.5)
                 + ... \\
                 &=& (29.1 \pm 2.5\pm 7.0) + ...~\hbox{eV} \\
                 &>& (29.1\pm 2.5\pm 7.0)~\hbox{eV},
\end{eqnarray*}
where the first errors are from fit and the second errors are the
systematic errors with the uncertainties from different fit
scenarios discussed above included. That is, the lowest value of
the $\Gamma_{e^+e^-}$ of the \Y\ is around 30~eV. This lower limit
on the leptonic partial width of the \Y\ is close to the
prediction from LQCD for a hybrid vector charmonium
state~\cite{chenying}.

On the other hand, if we take the Solutions with the largest
$\BR\times \Gamma_{e^+e^-}$ in Table~\ref{ddpi-table}, we obtain
$\Gamma_{e^+e^-}=(202\pm 13\pm 23)+...$~eV. This means that the
leptonic partial width of the \Y\ can be as large as 200~eV or
even higher based on current information, to be compared with the
predicted upper limit of 500~eV from the molecular
scenario~\cite{mole}. The other combinations of the Solutions
result in $\Gamma_{e^+e^-}$ values between $(30+...)$ and
$(200+...)$~eV.

If we assume these modes saturate the \Y\ decays, we determine the
\Y\ decay branching fractions to the above four modes. For the
most interesting mode, $\Y\to \pi\pi J/\psi$, we obtain
$\BR_{\pi\pi J/\psi}=(12.9\pm 1.3\pm3.9)\%$ (or a partial width of
$(7.2\pm 0.8\pm 2.2)$~MeV) for the case of smallest $\BR\times
\Gamma_{e^+e^-}$; and $\BR_{\pi\pi J/\psi}=(11.2\pm 1.1\pm1.9)\%$
(or a partial width of $(6.3\pm 0.7\pm 1.1)$~MeV) for the case of
highest $\BR\times \Gamma_{e^+e^-}$. In these two particular
cases, the branching faction of the $\Y\to \pi\pi J/\psi$ is very
big. We may also calculate the $\BR_{\pi\pi J/\psi}$ in the most
extreme case, i.e., taking the smallest $\BR\times
\Gamma_{e^+e^-}$ for $\pi\pi J/\psi$ and largest $\BR\times
\Gamma_{e^+e^-}$ for other modes, we find $\BR_{\pi\pi
J/\psi}=(2.1\pm 0.3\pm 0.7)\%$ (or a partial width of $(1.2\pm
0.2\pm 0.4)$~MeV).

However, the assumption that the $\omega \chi_{c0}$, $\pphc$,
$\pi^+ \pi^- J/\psi$, and $\ddpi$ modes saturate the \Y\ decays
may not be true. Being well above the thresholds of many final
states with $\eta_c$, such as $\pi\rho\eta_c$, $\omega\eta_c$, and
$\phi\eta_c$, and final states like $\eta h_c$, $\pi\pi\psi(2S)$,
and $K\bar{K}J/\psi$, \Y\ may decay into such final states with
substantial rates. In addition, the decays into open charm final
states other than $D \bar{D^*}\pi$ such as $D\bar{D}$,
$D\bar{D^*}+c.c.$, $D^*\bar{D^*}$, $D_s^+D_s^-$,
$D_s^+D_s^{*-}+c.c.$ are also possible, although the charmed
mesons are in relative P-wave. The \Y\ is very close to the
$D_s^{*+}D_s^{*-}$ threshold, the possible coupling to this mode
should also be investigated. Further information on these final
states will be important for a deeper understanding of the nature
of the \Y.

\acknowledgments

This work is supported in part by National Natural Science
Foundation of China (NSFC) under contract Nos. 11575017, 11235011,
11475187, and 11521505; the Ministry of Science and Technology of
China under Contract No. 2015CB856701; the CAS funding under
contract No. Y629132; Key Research Program of Frontier Sciences,
CAS, Grant No. QYZDJ-SSW-SLH011; and the CAS Center for Excellence
in Particle Physics (CCEPP).

\end{document}